\UseRawInputEncoding
\documentclass{article}

\usepackage[english]{babel}
\usepackage{latexsym}
\usepackage{comment}
\usepackage{mathtools, enumerate}

\usepackage[letterpaper,top=2cm,bottom=2cm,left=3cm,right=3cm,marginparwidth=1.75cm]{geometry}

\usepackage{amsmath}
\usepackage{graphicx}
\usepackage[colorlinks=true, allcolors=blue]{hyperref}
\usepackage{todonotes}

\date{}

\title{Abnormal light propagation and the underdetermination of theory by evidence in astrophysics}
\author{Felipe A.~Asenjo\footnote{felipe.asenjo@uai.cl}, Sergio A.~Hojman\footnote{sergio.hojman@uai.cl}, Niels Linnemann\footnote{niels.linnemann@unige.ch}, and James Read\footnote{james.read@philosophy.ox.ac.uk}}

\begin{document}
\maketitle

\begin{abstract}
We investigate the propagation of certain non-plane wave solutions to Maxwell's equations in both flat and curved spacetimes. We find that such solutions (or rather parts of them) exhibit accelerative behaviour, and in particular do not propagate on straight lines. Having established these results, we then turn to their conceptual significance---which, in brief, we take to be the following: (i) one should not assume that the part of electromagnetic waves from outer space that is subject to detection is localised onto null trajectories; therefore (ii) astrophysicists and cosmologists should at least be wary about making such assumptions in their inferences from obtained data, for to do so may lead to incorrect inferences regarding the nature of our universe.
\end{abstract}


\tableofcontents

\section{Introduction}

Einstein's 1905 theory of special relativity, as presented in \cite{Einstein1905}, involves several key inputs: the relativity principle (that the laws of physics are the same in all coordinate systems in uniform translatory motion with respect to one another), the light postulate (that the speed of light is independent of the speed of the source), the `standard' Einstein-Poincar\'{e} clock synchrony convention, and the homogeneity and isotropy of space. From the relativity principle and light postulate, one can derive straightforwardly that the speed of light takes the same fixed value $c$ in all inertial coordinate systems. So, in this sense, the constancy of the speed of light is close to being an input \emph{assumption} of the special theory \emph{circa} 1905---and, indeed, this constancy is often taken to be one of the defining features of the theory of relativity \emph{tout court}---both special and general.

Note, however, that the lesson of Einstein's 1905 paper was supposed to be the \emph{kinematical constraint} that all laws of physics are conditioned so as to be invariant under Lorentz transformations. Indeed, Einstein would in later life lament the special significance afforded to light and electromagnetism in his 1905 presentation of the theory, opting instead to cut directly to this conclusion:
\begin{quote}
The content of the restricted relativity theory can accordingly be summarised in one sentence: all natural laws must be so conditioned that they are covariant with respect to Lorentz transformations. \cite{Einstein1954}
\end{quote}
However, even if one accepts (with the later Einstein) the above kinematical constraint as the `essence' of special relativity (which also obtains locally in general relativity), one should avoid inferring too much from this constraint \emph{vis-\`{a}-vis} the nature of propagating relativistic fields. For example: although it is true that plane wave solutions to the Maxwell equations do indeed propagate with determinate velocity $c$, it is in fact far from obvious that this continues to be the case (a) when one considers other solutions to said equations, or (b) one accommodates for the possibility of spacetime curvature. Indeed, for over half a century relativistic causality has in fact been acknowledged to be a delicate business: see \cite{Butterfield} for a recent survey of this terrain, and \cite{AHbire,AH, LR,lightCP1,lightCP2,lightCP3,Mannheim2007, lightCP4,lastnewf1,lastnewf2,lastnewf3,lastnewf4} for particular investigations into the behaviour of wavelike solutions to the Maxwell equations in curved spacetimes.

For these reasons, it is arguably more foundationally careful to regard Einstein's 1905 theory of special relativity as taking the form of a \emph{self-consistent bootstrap} (cf.~\cite{Cao}): one begins with certain assumptions regarding the velocity of light; however, the conclusion at which one arrives (namely, the Lorentz invariance of the laws of physics) may---at least under certain circumstances---lead one to revise those very principles from which one began.

Our goal in the present article is to take up these issues, exploring in greater depth the propagation of certain non-plane wave solutions to Maxwell's equations (in particular Airy wave solutions) in both flat and conformally flat curved spacetimes. 
It is well-known that in a general curved spacetime (\emph{a fortiori} a flat or conformally flat spacetime) the Maxwell propagator takes support both on the lightcone and inside it
\cite{lightCP1,lightCP2,lightCP3,lightCP4}. However, the question remains whether this behaviour is the last word on the propagation of light. 
Here, we find that certain solutions to the Maxwell equations (or parts thereof) in conformally flat spacetimes can exhibit accelerative properties
(\emph{nota bene}: we will clarify different salient senses of wave velocity in what follows), and therefore that such wavepacket solutions { have \emph{parts} which do not propagate along null geodesics.  Note that this is consistent with the findings of \cite{AH,AHbire,lightCP4}, where it is shown that the \emph{whole} light packet moves along null geodesic in conformally flat spacetimes---the issue raised and focused on in this work is that some parts of the light wavepacket (usually the most intense ones) do not behave in such a manner.}\footnote{This is also in agreement with the fact that a finite-energy free-space electromagnetic
wavepacket, which is localized in three-dimensions, necessarily propagates at the speed of light, yet its net momentum is smaller than the energy divided by $c$ due to its spreading \cite{lekner2003}.}

Having established these results, we then turn to their conceptual significance---which, in brief, we take to be the following: 
(i) one should not assume that the part of electromagnetic waves from outer space that is subject to detection is localised onto null trajectories; therefore (ii) astrophysicists and cosmologists should at least be wary about making such assumptions in their inferences from obtained data, for to do so may lead to incorrect inferences regarding the nature of our universe.


The structure of this article is as follows. In \S\ref{s2}, we present the relevant background on waves and their velocities.
In \S\ref{s4}, we consider more generally non-plane wave solutions to the Maxwell equations in both Minkowski and cosmological FLRW spacetimes, taxonomising different ways in which non-null 
propagation of part of the wave might arise. In \S\ref{s5}, we focus in particular upon Airy solutions to the Maxwell equations in Minkowski and FLRW, and explore the propagation behaviour of these solutions. In \S\ref{s6}, we discuss the conceptual upshots of these results \emph{vis-\`{a}-vis}, among others, empirical underdetermination of light signals received in astronomy and cosmology and the theory-dependence of observation. 

\section{Notions of wave velocity}\label{s2}

In this section, we present the relevant background on (a) different notions of wave velocity (\S\ref{signalspeed}), and (b) velocities of non-plane waves (\S\ref{nonplane}).

\subsection{Background on wave velocities}\label{signalspeed}

Is it possible that electromagnetic waves \emph{in vacuo} might `signal' at some speed other than $c$? \emph{Prima facie}, there seem to be three distinct possible answers to this question:
\begin{enumerate}[(i)]
\item It is impossible that electromagnetic waves, { or parts of it,} could signal at some speed other than $c$.
\item Electromagnetic waves' signalling at some speed other than $c$ is possible in principle, but not in practice. This would raise a conundrum---possibly of deeper significance---as to why exactly we do not see such signals.
\item Electromagnetic waves' signalling at some speed other than $c$ is possible in practice. This, of course, would call for more research in this direction.
\end{enumerate}

Before we can make progress in deciding which of the above three answers is correct, however, we need first to get a better handle on what is meant by `signal', and what is meant by `signal speed', in the context of wave propagation. A modest working definition for `signal' is this: a propagating disturbance that encodes information where, standardly, the disturbance is taken to propagate locally. (The local nature of the propagation is an often tacitly assumed aspect which will become relevant later on.)

Disturbances can be modelled through wave packages. There are different ways to understand this:
\begin{enumerate}[(a)]
    \item The wave package \emph{per se} is taken to represent the disturbance.
    \item  Parts of the wave packages are taken to represent a notion of disturbance (say, the high-frequency components).
\end{enumerate}
Given this ambiguity, there are various distinct notions of signal speed which can be associated to a wave package, for example:
\begin{enumerate}[(I)]
\item The group velocity, which is the velocity with which the overall envelope of the wave's amplitudes propagates through space. (Regarding group velocity as the relevant notion of signal velocity would thereby be in line with (a) above.)
\item The front velocity, which is the velocity of the high frequency components of the wave (see e.g.~\cite{Shore}).
(Regarding the front velocity as the relevant notion of signal velocity would thereby be in line with (b) above.)
\end{enumerate}

Now, it is general consensus that (1) the concepts of phase velocity or group velocity are generally not valid concepts of signal velocity (although, of course, they can in cases serve as proxies for it, especially if they all agree), and that (2) the concept of front velocity---the high-frequency limit of the phase velocities of a wave package's components---serves as an upper bound to signal speed of a wave package, which, however, need not always be saturated (causal influences in water, for instance, propagate at speeds less than their front velocity): see \cite{GARRISON199819, Stenner}. 
Beyond this, however, the notion of the (signal) velocity of a wave is a delicate matter, and must be assessed on a case-by-case basis.

One other notion of wave velocity which will be of particular relevance in the context of our investigations in this article is the transverse `velocity' vector $\vec{v} = \vec{\nabla} \phi / \omega$, where $\phi$ is the phase of the wave and $\omega$ its frequency (this will be explained in more detail in \S\ref{s4.3})---such a notion of wave velocity is used in ongoing optics research \cite{Nichols, Nichols:22, PRL}, and has the merit of being associated with the normalised Poynting vector of an electromagnetic wave (and thus clearly of significance to energy propagation via the wave): for the exact chain of relations, see \cite[p.~4526]{Nichols} (also \cite{PRL, Petruccelli:13}). We will, indeed, have occasion to compute this quantity in later sections of this article.

\subsection{Velocities beyond isotropic waves}\label{nonplane}

Standardly, the aforementioned wave velocities are defined only with respect to plane waves or spherical waves. This, however, commits one to a basic ontology of waves whose fronts fulfil certain standards of isotropy; one might wonder as to the extent to which such notions of wave velocities can be extended beyond the plane or spherical wave in particular. To make progress in answering this question, recall first that a plane wave $\psi (\vec{x} ,t)$ can be written in the form
\begin{equation}
\psi(\vec{x}, t) = A  (\vec{k}) \exp ( \vec{k} \cdot \vec {x} - \omega t ),
\end{equation}
where $\vec{k}$ is a constant three-vector known as the `wave number', and $\omega$ is again constant, and denotes the frequency of the wave; moreover, the amplitude $A$ of the wave is constant.

One might, however, seek to generalise \emph{beyond} the notion of a plane wave. For example, consider generalised waves of the form
\begin{equation}\label{genwave}
\chi(\vec{x}, t) = A  (\vec{k}, \omega) \exp ( \vec{k} (\vec{x}, t) \cdot \vec {x} - \omega (\vec{x},t) t );
\end{equation}
in this case, the wave number and frequency are  spacetime-dependent quantities; the amplitude---which is a function of wave vector and frequency---is implicitly dependent on spacetime quantities through its dependence upon $\vec{k}(\vec{x},t)$ and $\omega(\vec{x},t)$. 
One might object that such a liberalisation of the notions of wave vector and frequency goes against the spirit of introducing the wave vector and frequency to begin with; after all, on a plane wave approach, wave vector and frequency parameterise a specific plane wave and thereby allow for the packaging of specific plane waves in a weighted manner through a wave vector- and frequency-dependent amplitude. However, even if each wave vector is spacetime-dependent, it is not the case that different waves cannot be summed; in this sense, the generalisation \eqref{genwave} is perfectly legitimate.
Needless to say, the generalised waves could be decomposed further into plane waves proper. But the point of the general wave form is to allow for the setting up of notions of wave velocity that are not  limited to plane waves \emph{specifically}, independent of the mathematical fact that (as is of course well-known) plane waves form a basis for the solution space for a wave equation.

\begin{figure}
    \centering
    \includegraphics[scale=0.3]{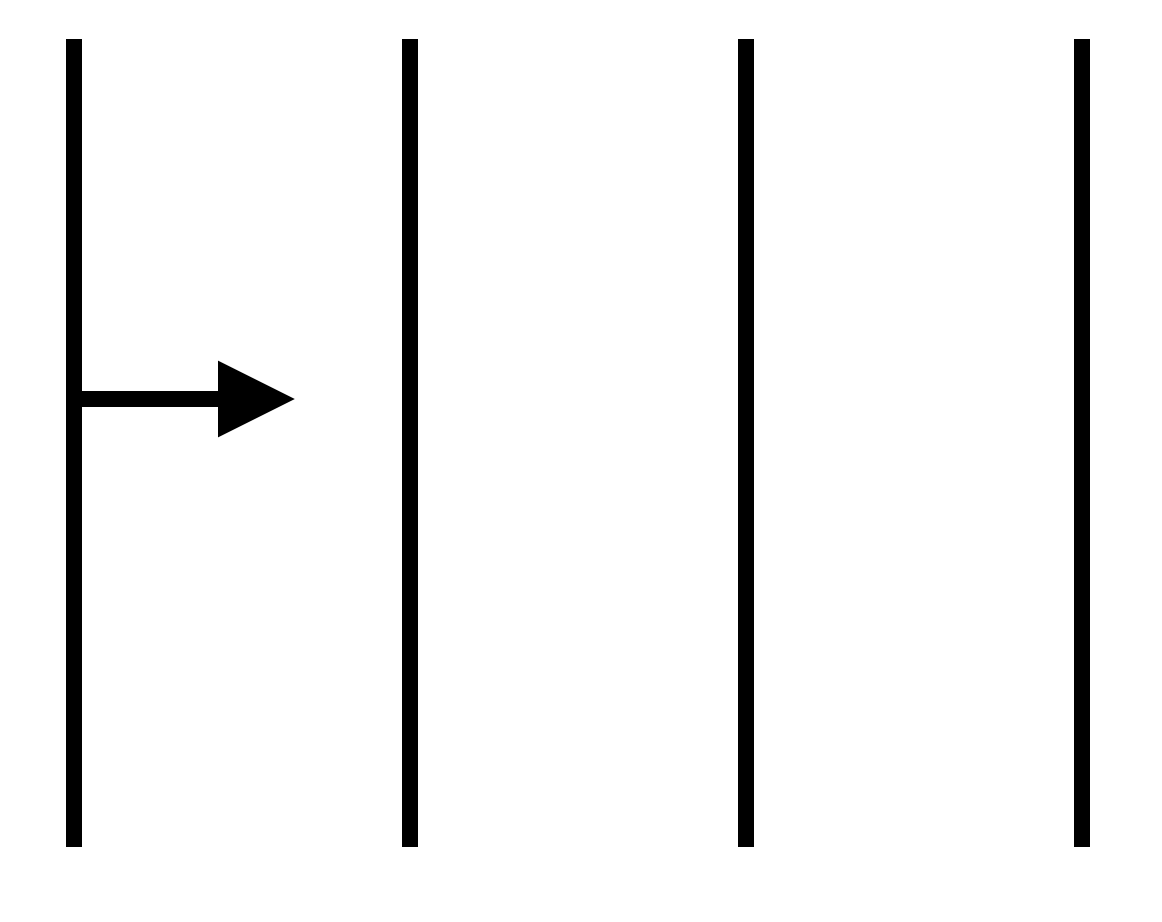}
    \caption{Wave fronts for a plane wave.}
    \label{fig:regular}
\end{figure}

\begin{figure}
    \centering
    \includegraphics[scale=0.3]{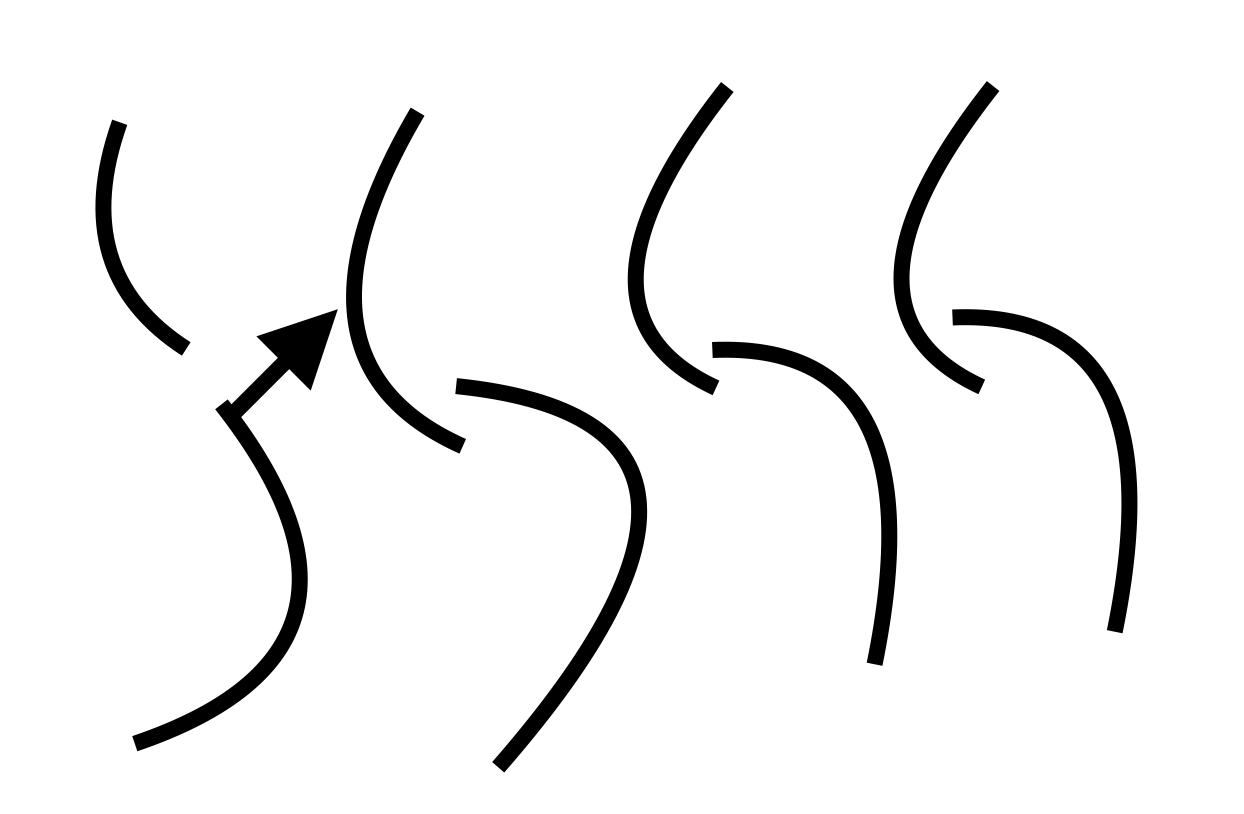}
    \caption{Discontinuous wave fronts for a generalised `wave'.}
    \label{fig:discontinuous}
\end{figure}

\begin{figure}
    \centering
    \includegraphics[scale=0.3]{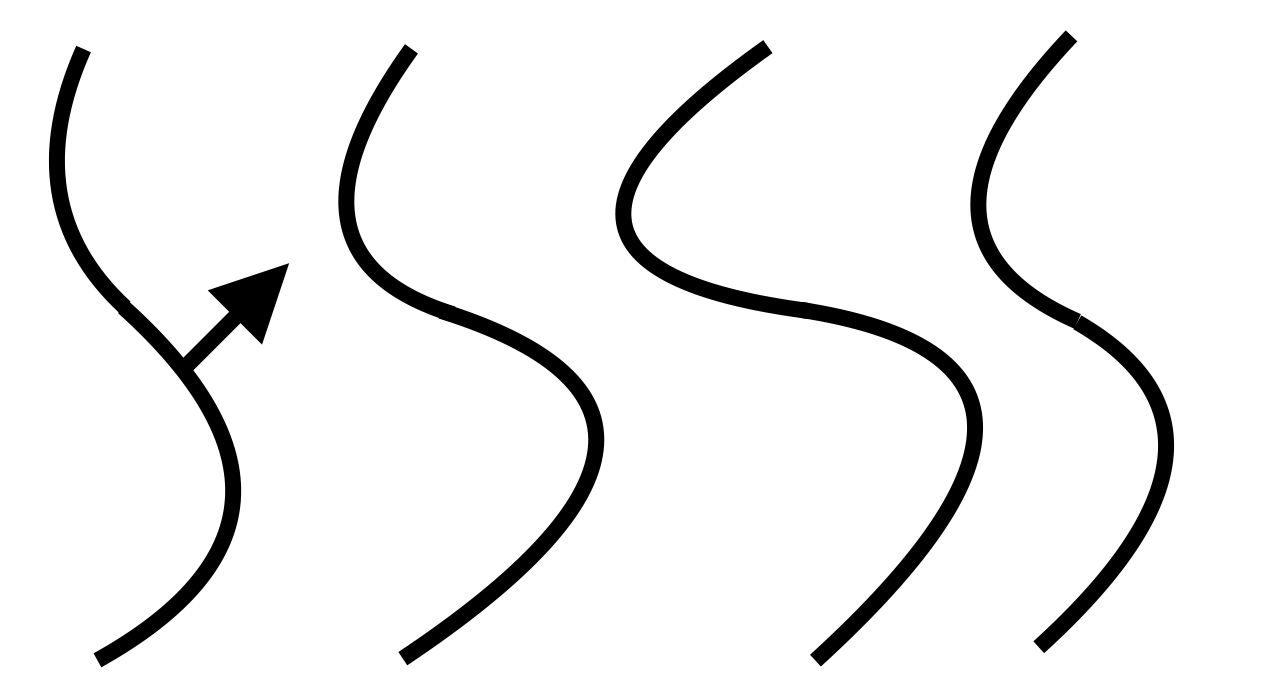}
    \caption{Continuous but spacetime-dependent wave fronts for a generalised wave.}
    \label{fig:curvy}
\end{figure}

Physically, the wave number picks out level surfaces in space which correspond to wave crests: for the canonical plane wave, these are evenly-spaced, as per Figure \ref{fig:regular}. In the more general case in which the wave number is an arbitrary function of spacetime, however, it may manifest discontinuities and caustics, as in Figure \ref{fig:discontinuous}. In that case, it is by no means clear that notions such as phase, group, or front velocity will be well-defined mathematically. Therefore, if one is to continue to make sense of such notions of wave velocities once one has generalised beyond the plane wave in the above manner, it is prudent to impose minimal restructions upon the wave number---in particular, it is prudent to impose that its level surfaces be \emph{continuous} (or maybe even smooth) in space. In this way, the wave vector can nevertheless pick out `non-plane wave' surfaces, such as those of Figure \ref{fig:curvy}.
In a like manner, in order for notions of phase, group, and front velocity to be well-defined in the general case, one of course will also require that the wave frequency $\omega$ be continuous, for essentially the same reasons as above.


\section{Maxwell equations in flat spacetime and FLRW cosmology}\label{s4}

In this section, we present two solutions for abnormal light propagation arising from the vacuum Maxwell equations in flat spacetime and in  Friedman-Lemaître-Robertson-Walker (FLRW) cosmology. 
In particular, we show that this abnormal light propagation involves beam constituents which move on bent rather than straight trajectories---and so not on null geodesics. These can occur for two different reasons: (i) the existence of second-order amplitude wave gradients (giving rise to the known `Bohm potential effects'), and (ii) the existence of second-order polarization wave gradients. Both are discussed below, with corresponding particular solutions.

The structure of the section is this. In \S\ref{s4.1}, we remind the reader of the relevant details of the Maxwell equations in curved spacetimes. In \S\ref{s4.2}, we consider amplitude gradient effects. In \S\ref{s4.3}, we consider polarisation gradient effects.

\subsection{Maxwell equations in flat and curved spacetimes}\label{s4.1}

In general, the Maxwell equations in vacuum flat spacetime for the electric field ${\bf E}$ and the magnetic field ${\bf B}$,
are written (in vectorial notation) as
\begin{eqnarray}
\nabla\cdot {\bf B}&=&0\, , \quad \nabla\cdot {\bf E}=0\, ,\nonumber\\
\partial_t {\bf E}&=&\nabla\times {\bf B}\, , \quad \partial_t {\bf B}=-\nabla\times {\bf E}\,  .
\label{Maxflat}
\end{eqnarray}
The abnormal propagation of light in this context (i.e.,~\emph{even in flat spacetime}) has been studied theoretically and experimentally as solutions of these equations for electromagnetic waves (see \emph{inter alia} \cite{siviloglou2007,siviloglou2008,bekenstein2014shape,beenstein2012prl,patsyk2018observation,Hacyan2011,Parinaz2016,Chremmos2013,Siviloglou2007a,chong2010,peng2016,ping2013}). 

Before we consider this in more detail, however, let us first demonstrate that in FLRW cosmologies, electromagnetic wave propagation can in fact be described by a similar set of Maxwell equations. As is well-known, the Maxwell equations in curved spacetime can be written
\begin{equation}
    \nabla_\mu F^{\mu\nu}=0\, , \qquad \nabla_\mu F^{*\mu\nu}=0\, ,
\end{equation}
in terms of compatible covariant derivative operators for a general spacetime metric $g_{\mu\nu}$.
As usual, we can now define the vectorial electric and magnetic fields as
$E_i=F_{i0}$, $D^i=\sqrt{-g} F^{0i}$, $B^i=\varepsilon^{0ijk}F_{jk}$ and $\varepsilon^{0ijk}H_k=\sqrt{-g} F^{ij}$, with relations
\begin{eqnarray}
D^i&=&-\sqrt{-g}\frac{g^{ij}}{g^{00}}E_j+\varepsilon^{0ijk}\frac{g_{0j}}{g_{00}}H_k\, ,\nonumber\\
B^i&=&-\sqrt{-g}\frac{g^{ij}}{g^{00}}H_j-\varepsilon^{0ijk}\frac{g_{0j}}{g_{00}}E_k\, .
\label{relationgeneralelectrimagneti}
\end{eqnarray}
Then, the Maxwell equations can be written
\begin{eqnarray}
\nabla\cdot {\bf B}&=&0\, ,\quad \nabla\cdot {\bf D}=0\, ,\nonumber\\
\partial_0 {\bf D}&=&\nabla\times {\bf H}\, , \quad \partial_0 {\bf B}=-\nabla\times {\bf E}\, .
\label{Max1}
\end{eqnarray}
The above system is general for any spacetime. However, in what follows let us now consider specifically the FLRW metric 
\begin{equation}
    g_{\mu\nu}=(-1,a^2,a^2,a^2)\, ,
\end{equation}
where $a = a(t)$ is the usual FLRW scale factor. In this case, relations \eqref{relationgeneralelectrimagneti} reduce to
\begin{eqnarray}
{\bf D}&=&a {\bf E}\, ,\quad {\bf B}=a {\bf H}\, ,
\end{eqnarray}
and then, the above Maxwell equations \eqref{Max1} become
\begin{eqnarray}
\nabla\cdot {\bf B}&=&0\, ,\quad 
\nabla\cdot {\bf D}=0\, , \nonumber\\
\partial_\tau {\bf D}&=&\nabla\times {\bf B}\, , \quad
\partial_\tau {\bf B}=-\nabla\times {\bf D}\, , 
\label{Max2}
\end{eqnarray}
---i.e.,~just like the free Maxwell equations in flat spacetime!---by defining the conformal time
\begin{equation}
    \tau=\int \frac{dt}{a}.
\end{equation}
Therefore, everything that is fulfilled for electromagnetic waves in free flat space occurs also for cosmological FLRW electromagnetic waves, in the conformal time $\tau$. In flat spacetime, $a=1$ and $\tau=t$. In a sense, of course, this is unsurprising, since (a) the FLRW solution is conformally flat, and (b) the source-free Maxwell equations are conformally invariant, so any solution of said equations in Minkowski spacetime will remain a solution in FLRW spacetime. 


With the set of equations \eqref{Max2} in hand, we find that the magnetic field ${\bf B}$ follows the general wave equation
\begin{equation}
    \left(\partial_\tau^2-\nabla^2\right){\bf B}=0\, .
    \label{waveeq}
\end{equation}
(A similar equation can be found for the field ${\bf D}$---we discuss this further below.)
Usually, \eqref{waveeq} is solved in terms of plane waves travelling to the speed of light (or, in relativistic vocabulary, on null geodesics), as well as superpositions thereof. However, this does not necessarily allow one to appreciate how abnormal the light behaviour can be for some of those superposed solutions. As an exemplification of this, in the following two subsections we obtain two class of solutions to the Maxwell equations (without making any approximation)---in the first case, the solutions involve second-order amplitude gradients; in the second, they involve second-order polarisation wave gradients. Note, to repeat, that both of these classes of solutions (for the reasons articulated above) are solutions both of the Maxwell equations in flat Minkowski spacetime \emph{and} in the (curved) FLRW spacetime which we take to represent accurately the large-scale structure of our universe.

\subsection{Amplitude gradient effects}\label{s4.2}

When the wavepacket solution for electromagnetic wave propagation is allowed to have amplitude gradients, then it modifies the propagation of its parts. This has been reported extensively in the context of the paraxial and non-paraxial propagation of light in both flat and curved spacetime geometries \cite{bekenstein2014shape, patsyk2018observation,beenstein2012prl}; in these spacetimes, it has been shown that such wavepackets are composed by different parts following non-geodesic trajectories, and which indeed follow curved trajectories due to the interaction of amplitude properties of the wave with its medium. Here, we show explicitly (and arguably for the first time) that such behaviour holds in flat and conformally flat spacetimes without any approximation (such as the paraxial approximation).


In this subsection, we present the simplest case of such effects for a cosmological scenario; in particular, we show that an electromagnetic field is described by a wavepacket with different lobes propagating in an accelerated fashion due to amplitude gradient effects.
We look for a solution in the form ${\bf B}=B(u,y) \exp(i v)\hat z$, with only one component (with $u=x-\tau$ and $v=x+\tau$, the light-cone coordinates). In this case, the wave equation \eqref{waveeq} 
becomes
\begin{equation}
    \left(4 i\partial_u +\partial_y^2 \right)B=0\, .
\end{equation}
This is a Schr\"odinger-like equation, which has several known solutions. One of them is in the form of an Airy function \cite{berry1979,lekner2009}. Therefore, in our case, the field can propagate as
\begin{equation}
    B(\tau,x,y)=B(u,y)=\mbox{Ai} \left(2y-u^2+i \beta u \right)\exp\left(\beta y-\beta u^2+i\frac{\beta^2}{4}u+2i u y-\frac{2}{3}iu^3 \right)\, ,
    \label{Airygensolution}
\end{equation}
where $\mbox{Ai}$ is the Airy function, and $\beta$ is an arbitrary parameter that allows us to model a finite-energy wavepacket \cite{lekner2009}. 

\begin{figure}
    \centering    \includegraphics[scale=0.7]{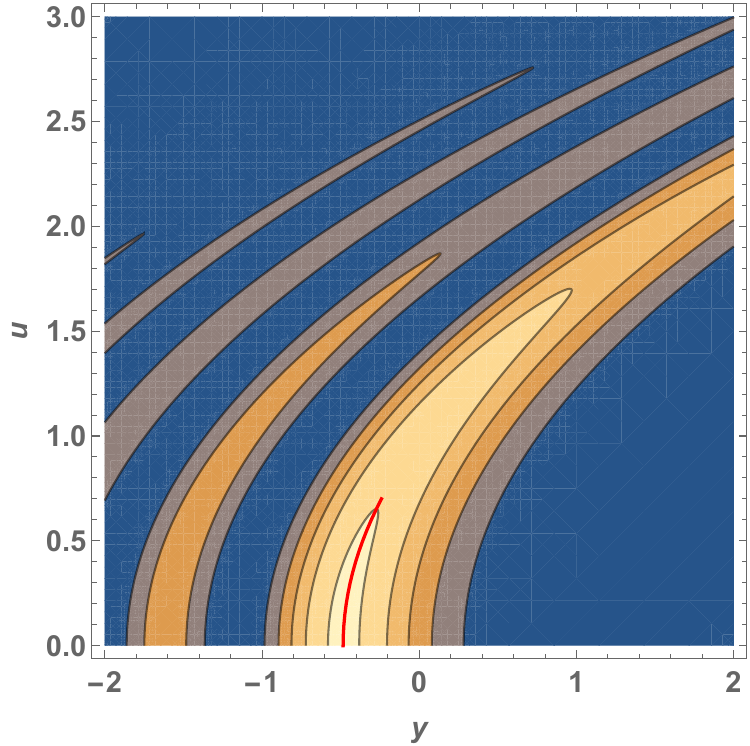}
    \caption{Contour plot, in the $y$-$u$ plane with $\beta=0.1$, of the square of intensity of wavepacket \eqref{intensityb} for the different lobes of the solution. The red line shows the deflection of this maximum intensity of this solution on this space.}
    \label{fig:my_label}
\end{figure}

The above solution presents different intensity lobes that follow curved trajectories in the $y$-$u$ plane, i.e., as parts of the wavepacket propagate along the light-cone direction $u=x-\tau$, it deflects also on transversal $y$-direction. This behaviour does not occur for a plane wave, which does not manifest any kind of deflection in its transverse direction. This can be studied numerically through the intensity of this wavepacket
\begin{eqnarray}
b=\sqrt{{\bf B}^* {\bf B}}=\sqrt{B^* B}\, .
\label{intensityb}
\end{eqnarray}
In Figure \ref{fig:my_label}, we contour plot the square of intensity $b^2$, on the $y$-$u$ plane with $\beta=0.1$, to show that this Airy solution has several lobes that behave differently, deflecting in a different manner. In the plot, the red line shows the curved trajectory of the maximum intensity of this wavepacket. In the usual plane wave solution, all trajectories should be vertical lines in such plots. Therefore, the abnormally-propagating solution 
\eqref{Airygensolution} deflects itself during its propagation due to its dynamics on the transverse plane.

Similarly, we can calculate the field ${\bf D}$ from \eqref{Max2}---this is important, of course, in order to check that we have a genuine solution to the vacuum Maxwell equations. For this case, ${\bf D}=\nabla\times(f\hat z)$, where $\partial_\tau f=B \exp(i v)$. Thus,
\begin{equation}
f=f(\tau, x,y)=f_1\,  \mbox{Ai}(\zeta) +f_2\,  d\mbox{Ai}(\zeta)/d\zeta,
\end{equation}
where
\begin{align}
\zeta&=2y-u^2-i \beta  u, \\
f_1&=-\zeta^2/2+\int d\tau \exp(\beta y-\beta u^2-i({\beta^2}/{4})u-2i u y+({2}/{3})iu^3+iv), \\
f_2&=-\int d\tau \partial_\tau\zeta f_1.
\end{align}
In the typical case of plane waves (with constant amplitudes), the only non-zero component of the vector field ${\bf D}$ is in the $y$-direction, and the flux of energy (i.e.,~the Poynting vector) is in the $x$-direction. However, in this case,
the vector field ${\bf D}$ has two components, 
${\bf D}=\partial_y f\hat x-\partial_x f\hat y$.
Thus, the Poynting vector ${\bf S}=({1}/{2})({\bf D}^*\times {\bf B}+{\bf D}\times {\bf B}^*)$ has components
\begin{align}
S_x &= -\frac{1}{2} \left(\partial_x f^* B e^{iv}+\partial_x f B^* e^{-iv}\right), \\ 
S_y &= -\frac{1}{2} \left(\partial_y f^* B e^{iv}+\partial_y f B^* e^{-iv}\right).
\end{align}
 Therefore, in this wave solution the energy flux follows a curved trajectory in the $y$-$u$ plane. 
This is due only to the transversal (extended) spatial dependence of the amplitude of the wave, which now depends also on the $y$-direction.

{
To stress, these Airy-type solutions are of course expressible as a combination of plane waves, i.e.~their
Fourier transforms exist and  can be calculated \cite{Siviloglou2007a}. The `average behaviour' of such solutions in the sense of the Ehrenfest
theorem is the `normal' one, but the Airy function main lobe exhibits acceleration which is observable experimentally, while the rest of the (dimmer) lobes accelerate in the opposite direction.

}

{\color{black}

It is worth noting also that there can  be non-plane wave solutions of abnormal nature every part of which still fully propagates along the light-cone (i.e., along the light-cone coordinates $u$ and $v$) \cite{tetryakov}.
To see this, make the following ansatz to the solution of \eqref{waveeq}:
\begin{eqnarray}
    {\bf B}(t,x,y)=b(u,v)\exp(i k y)\hat z\, 
\end{eqnarray}
 which gives
\begin{equation}
    4\frac{\partial^2 b}{\partial u\partial v}=k^2 b\, .
    \label{waveequv}
\end{equation}
Remarkably, this equation is not only solvable by plane waves: one solution due to \cite{tetryakov} is
\begin{equation}
    b(u,v)={\mbox{Ai}}\left(\frac{k^2}{4 \delta^2}u+\delta\sqrt{v} \right){\mbox{Ai}}\left(\frac{k^2}{4 \delta^2}u-\delta\sqrt{v} \right)\, .
\end{equation}
This solution is not separable in $u$-$v$ variables, and its behaviour depends on the uneven interaction of the propagation along $u$ and $v$ coordinates, where $\delta\neq 0$ is an arbitrary constant.

As before, we can calculate the Poynting vector of this wavepacket.
In this case, the Poynting
vector is along the propagation direction ${\bf S}=S\hat x$, where 
\begin{eqnarray}
    S=-b\int dt\, \frac{\partial b}{\partial x}\, .
\end{eqnarray}
Then, the flux of energy of this electromagnetic wave is along its propagation direction, as  {\color{black}it is for} plane waves, with no deviation on its trajectory. However, this flux of energy is not constant along the path of the wave, so   {\color{black} this} behaviour is very different from that of plane waves.

}

\subsection{Polarisation gradient effects}\label{s4.3}

To complement the above-discussed situations in which the propagation of electromagnetic waves can be affected by an amplitude gradient, we ask in this subsection whether effects on \emph{polarisation} can likewise induce abnormal wave propagation. In general, the wave equation \eqref{waveeq} describes the behaviour of polarised waves, and therefore more general solutions than the plane wave can again be identified.
For example, recently it has been shown (both theoretically and experimentally) in  \cite{Nichols:22} for the Eikonal transport equation approximation that second-order gradients on the phase of polarised waves in vacuum flat spacetimes can also induce beam deflections on the wave.
Here we demonstrate that such effects occur already without any approximations to the Maxwell equations. 

Consider a magnetic field with a transverse polarized part in the form
\begin{eqnarray}
{\bf B}(\tau,x,y,z)=\rho^{1/2} \left[ \cos\gamma\,   \hat x+\sin\gamma\,   \hat y\right] \exp(i \phi +i \omega \tau)+B_z \hat z\, ,  
\label{ansatzgrad}\end{eqnarray}
where $\rho$ is the amplitude of the wave,  $\gamma$ is the polarization angle,  $\phi$ is total wave phase, and $\omega$ is the wave frequency. The physical quantities $\rho$, $\gamma$, and $\phi$ can have full dependence on (conformal) time and space coordinates. The frequency $\omega$ is a constant for this solution. $B_z$ denotes a longitudinal part of the field.
Using this ansatz in \eqref{waveeq} for the transverse part, we obtain the four equations
\begin{eqnarray}
    \nabla\cdot \left(\rho \nabla\gamma\right)&=&0\, ,\nonumber\\
    \nabla\gamma\cdot \nabla\phi&=&0\, ,\nonumber\\
    \nabla\cdot \left(\rho \nabla\phi\right)&=&0\, ,\nonumber\\
    \left|\nabla\phi \right|^2-\omega^2-\frac{\nabla^2\sqrt{\rho}}{\sqrt{\rho}}+\left| \nabla\gamma\right|^2&=&0\, .
    \label{polarizagradientequations}
\end{eqnarray}
On the other hand, $B_z$ must fulfil
\eqref{waveeq}.
In principle, we see already from this (in particular from the last equation) that the gradient of the polarization angle can induce a deflection on the trajectory of the wave, calculated through the second order gradient of the phase $\nabla(\left| \nabla \phi\right|^2)$ \cite{Nichols:22}.

A particularly neat example of this effect can be obtained for a perturbative solution. Let us suppose that
\begin{eqnarray}\label{eq:phi}
    \rho (z)&\approx& \rho_0+2\epsilon\, \rho_1 (z)\, ,\nonumber\\
    \gamma(x,y)&\approx& \epsilon^{1/2}\, \gamma_1(x,y)\, ,\nonumber\\
    \phi(x,y,z)&\approx& \omega\, z+\epsilon\, \phi_1(x,y,z)\, ,
\end{eqnarray}
 where $\epsilon$ is the small parameter that we use to keep to track of the perturbative order in the equation. Here $\rho_1$, $\gamma_1$, and $\phi_1$ are the perturbations, while $\rho_0$ is a constant. Thereby, the transverse part of this solution is a polarised perturbation of a plane wave travelling in the $z$-direction. 

The first and second equations of set \eqref{polarizagradientequations} (both at $\mathcal{O} (\epsilon^{1/2}$)) allow us to establish that
\begin{equation}
    \gamma_1(x,y)=\omega \alpha x y\, ,
\end{equation}
where $\alpha$ is a constant. In addition, the third and fourth equations (both at $\mathcal{O} (\epsilon)$)) allow us to find that
\begin{eqnarray}\label{eq:phi1}
\rho_1(z)&=&\frac{\alpha^2}{2}z^2\, ,\nonumber\\
    \phi_1(x,y,z)&=&\frac{\alpha^2}{2\omega}z\left(1-\omega^2(x^2+y^2) \right)\, .
\end{eqnarray}
Finally, from constraint \eqref{Max2}, the above transverse solution for the ansatz allows us to find $B_z$ at order $\mathcal{O} (\epsilon^{1/2})$. This is
\begin{equation}
    B_z(\tau, x, z)=i \epsilon^{1/2} \sqrt{\rho_0}\,\alpha\, x \exp(i\omega z+i\omega\tau)\, ,
\end{equation}
which also solves \eqref{waveeq}  identically.\footnote{We are not aware of an example which fully isolates the polarisation gradient effect from the amplitude gradient effect.}

From the above solution we can see that small changes on the polarization angle will induce small changes on the wave trajectory. 
The transverse velocity $\bf v$ can be obtained through the  gradient of its phase ${\bf v}=v_x\hat x+v_y\hat y+v_z\hat z=\nabla\phi/\omega$ \cite{Nichols:22} (on this notion of velocity, recall our discussion in \S\ref{signalspeed}). Thus, for example, using the phase $\phi$ from \eqref{eq:phi} as well as the specification for $\phi_1$ from \eqref{eq:phi1}, this wavepacket {\color{black} has a component $v_y=-\epsilon \alpha^2 y z $, which implies}
\begin{equation}
    \left|\frac{d v_y}{dz}\right|=\epsilon \alpha^2 y\, .
    \label{gradientvelcoyz}
\end{equation}
{\color{black} Therefore,} the wave propagates longitudinally in $z$-direction,  {\color{black}deflecting} along the transverse $y$-direction due to the polarization gradient effect. Similar behaviour happens for the $x$-direction.

As before, this can be studied using the Poynting vector of this solution.  
Its complete calculation is highly involved and not very illuminating, but instead one can obtain the different $\epsilon$ order corrections for the solution. For the case in hand, the Poynting vector values are ${\bf S}=\frac{1}{2}({\bf D}^*\times {\bf B}+{\bf D}\times {\bf B}^*)=S_x \hat x+S_y\hat y+S_z\hat z$, where
\begin{eqnarray}
    S_x&=&\mathcal{O} (\epsilon^{3/2})\, ,\nonumber\\ S_y&=& \epsilon \alpha^2  y z \rho_0 +\mathcal{O} (\epsilon^{3/2})\nonumber\\
    S_z&=&-\rho_0+\epsilon^{1/2}\frac{\alpha\rho_0}{\omega}+\epsilon \frac{\alpha^2}{2}\left( \rho_0(x^2+y^2)-2 z^2-\rho_0/\omega^2\right) +\mathcal{O} (\epsilon^{3/2})\, ,
\end{eqnarray}
implying that the main flux of energy occurs in the $z$-direction (which is the expected result as it is the non-perturbative plane wave solution), along with several corrections. However, we find that, at $\epsilon$ order, energy also flows in the $y$-direction. Thus, while the wave mainly propagates in $z$-direction, it also deflects in $y$-direction. This is in agreement with the velocity gradient \eqref{gradientvelcoyz}, implying that {\color{black} $|v_y|=S_y/\rho_0$, can be related to} the normalised Poynting vector of this electromagnetic wave.

\section{The case of Airy beams}\label{s5}


So far, we have seen that both amplitude and polarisation gradient effects can affect the propagation trajectories of electromagnetic waves in both flat and FLRW spacetimes. In this section, we explore the physical status of optical Airy beams, the discussion of which will also help us to understand better the status of the specific Airy-type solutions in curved spacetimes discussed in the previous section. Generally from now on, we will use Airy beam scenarios as paradigmatic cases for abnormal light propagation (as by now familiar from wave propagation in dielectric media), but this focus is in no way intended to suggest that abnormal propagation is limited only to Airy beam behaviour.

Airy beams in the experimental literature are usually associated to solutions of the paraxial approximation of the Helmholtz equation. (This is in contrast to the case discussed above, which although (a) written in terms of Airy functions, and (b) exhibiting an `accelerating' central lobe, is in fact an \emph{exact} solution to the vacuum Maxwell equations.) That is, the Helmholtz equation $\nabla^2 f + k f = 0$ with ansatz $A(r) = u(r) \exp(ikz)$ is approximable to 
    \begin{equation}
    \nabla_{\perp}^2 u + 2 i k \frac{\partial u}{\partial z} = 0
    \end{equation}
    provided that $|\frac{\partial^u}{\partial z^2}| \ll |k \frac{\partial u}{\partial z}|$.\footnote{Notably, the equation is mathematically equivalent to the Schr\"{o}dinger equation---thus the appearance of Airy waves in quantum mechanics too (albeit there they are `over time and space' and not just space).}
Apparent core features of Airy beams include that they are weakly diffracting (amplitude profile stays close to invariant), curve in propagation (`parabolic shape') and can be self-healing (the profile can recover its original form when obstacles are placed in the path of their main lobe). All these three aspects of behaviour can be explained by way of a detailed analysis of Airy beams in terms of a ray decomposition approach: see \cite{Kaganovsky}. Interestingly, Airy beam solutions to the paraxial approximation of the Helmholtz equation are the only diffraction-free solutions in one transverse-dimension. This, however, leaves open the possibility for non-diffraction-free paraxial accelerating waves or those with more than one transverse direction.

What about the non-paraxial case? Consider the 2d case:
    \begin{equation}
    \frac{\partial^2 f}{\partial{x^2}} + \frac{\partial^2 f}{\partial{z^2}} + k^2 f = 0,
    \end{equation}
  As reviewed in \cite{Airyreview}, so-called half-Bessel beams are diffraction-free solutions to the 2d Helmholtz equations. Notably, they can follow circular, ellipsoidal and other paths, whereas the Airy beams---solutions to the paraxial approximation of the 2d Helmholtz equation---trace out parabolic paths.

What then is the significance of Airy beams \emph{qua} physical signals? The work of \cite{Kaganovsky}, which analyses Airy beams in terms of geometric rays, is especially helpful in answering this question. The set-up is as follows: the non-paraxial equation is considered in two dimensions,
\begin{equation}
\partial_x^2 u + 2 i k \partial_z u = 0,
\end{equation}
together with an initial field distribution in the aperture plane $z=0$ of form $u (z=0) = \text{Ai} (x/x_0)$. The radiated field $f = u \exp(i k z)$ is then given by Kirchhoff-Huygens integral
\begin{equation}
U(r) = 2 \int_{\infty}^{\infty} dx' \phi_0 (x') \partial_{n'} G(r, r')
\end{equation}
with Rayleigh-Sommerfeld diffraction Green function $G$. One can distinguish between contributions to the field $U$ from the main lope of the aperture (call this $U_1$) and contributions from the remainder of the aperture (call this $U_2$). The analysis of \cite{Kaganovsky} then leads to the \emph{prima facie} surprising insight that the main curved lobe of the Airy beam ``is not generated by the first main lobe of the aperture field, but by some other part" (p.~2).  In particular, it is concluded in light of the analysis that an Airy beam is \textit{not} a beam in the sense of ``a local wave function whose evolution along its propagation trajectory is described by a local self-dynamics" (p.~2).

However, the results of \cite{Kaganovsky} need to be put into context.
It is the field profile of the beam, rather than the center of mass, that has immediate influence on optical processes.  As for instance stressed in the review \cite{Airyreview}, ``the majority of optical processes like nonlinear effects, \emph{detection}, imaging etc., are directly dictated by the field profile itself, not by the `center of mass' of the beam, which is a mean property (averaged over the beam cross section)." (p.~686, our emphasis).\footnote{On p.~687, the authors also write: ``The study of optical accelerating waves effectively began with the understanding that what really matters is not the trajectory of the `center of mass,' but the curved evolution of the field itself. Thus, a truncated Airy beam (which is physically realizable since the beam must have finite power), interacts with its surroundings---whether they are particles or other waves---according to its accelerating field structure, not according to its centroid, which moves instead on a straight line so as to ensure the preservation of transverse momentum."}  Generally speaking, as \cite{patsyk2018observation} discuss, the main feature of accelerating beams carrying finite power is that their main lobe follows a curved path while their center of mass does follow a straight trajectory. Again, this has significant consequences for light-matter interactions, as all such interactions depend on the local intensity of the beam rather than its center of mass; in fact, this feature has enabled a variety of applications, including curved plasma channels, manipulation of microparticles, laser micromachining, single-molecule imaging, and light-sheet microscopy, among others.

How do these Airy wave solutions here relate to the ones found in the previous section? We take much of the insights from the former types of Airy waves to carry over to the latter types of Airy waves given that they have qualitatively the same feature of a divergence between main lobe and side lobe propagation, leading to an `accelerative' property of the former.
Even then one might claim that the field profile of an Airy beam, \emph{qua} interference result from non-local sources, can hold only if specific coherence conditions are met.\footnote{We thank an anonymous referee for bringing this point to our attention.} At the same time, the Airy beam can be obtained straightforwardly in experimental set-ups by letting an apparently `local' Gaussian beam pass through a suitable filter. 
For instance, Airy beams, in the sense of \S \ref{s4.2}, can be implemented straightforwardly (given that the Fourier transform of the Airy wave is a Gaussian beam with an additional cubic phase)  by using a spatial light modulator \cite[p.~687]{Airyreview},\footnote{See also \cite{siviloglou2008, broky2008}.} specially-designed lenses \cite{yalizay2010,papazoglou2010} on a Gaussian laser mode, or plasmonic metalenses \cite{TellezLimon}; polarised beams in the sense of \S \ref{s4.3} can in principle be generated using an initial polarized light  beam that passes through spatial light modulators 
and polarization filters \cite{Nichols:22}.\footnote{We thank Zhigang Chen and Nikos Efremidis for correspondence on these issues.} Even more, Airy forms of electromagnetic propagation can be induced  by the reflection and transmission of electromagnetic waves in  plasmas with inhomogeneous densities \cite{jahnG}. As plasmas are ubiquitous in nature,  Airy forms of propagation---one might argue---are expected.

How does this fit together? Now, what is key in understanding the situation is to regard the relevant scales: (1) From far away, any electromagnetic propagating wave has to move on null geodesics (that follows from taking the geometric optics limit \cite[\S22.5]{MTW}); at this scale, the wave package has no internal structure. (2) At any scale, any electromagnetic solution can be analysed in terms of plane wave decomposition. If one does so for the Airy wave coming out of the Gaussian wave from the aperture of a specific filter, one realises that the Airy wave's main lobe is the result of interference from different parts of the aperture of the filter---but not from that part of the aperture which `looks' like its local origin on the aperture. So, we can say that we can only expect Airy waves to occur provided that the size of the aperture relative to that of the main lobe remains effectively large (simply as the main lobe results from contributions all over the aperture and not from its `local' past). Strictly speaking, then, all forms of underdetermination in astrophysics (see \S\ref{s6.1}) arising from Airy waves (and potentially other accelerated beams) appear under the assumption that light sources or optical filters are available in outer space which are sufficiently larger in extension than the main lobe's circumference.

We grant that it remains open whether there are physical realisers of such conditions in astrophysics---our point, however, is that such conditions need to be kept in mind. But there are also related questions to attend to here, namely: for how long would such beams persist? In \cite{broky2008self, zhuang2015quantitative}, it is shown in the laboratory setting that an Airy beam can travel for a distance of more than 50 times its diameter prior to losing its shape via diffraction. Assuming that the same ratio would hold in cosmological contexts, this perhaps undermines the claim that such beams could travel cosmological distances. However, even accepting this (and there remains more work to be done regarding whether other solutions could propagate for greater distances before losing shape via diffraction), this highlights that such possibilities need to be considered particularly in the context of astrophysics on comparatively short scales.

\section{Foundational consequences}\label{s6}

Up to this point, we have made the case that once one generalises beyond plane wave solutions to the Maxwell equations in both flat and conformally flat curved spacetimes, the propagation of the waves may be affected so as to lead to non-null motion of parts of such a structured wavepacket. (Of course, one expects the situation to become yet more complicated when one moves beyond the case of conformally flat spacetimes; that, however, is not the focus of the present article.) 

\emph{Prima facie}, this has the potential to be a profound result, given that the standard assumption in practising astrophysics and cosmology is that electromagnetic signals received at Earth travel on null geodesics, and are localised entities. In particular, given the cornucopia of different ways in which wave-like solutions to the Maxwell equations have been generated in the lab as well as brought into application, the assumption that the actually detected parts of propagating solutions to the Maxwell equations themselves resemble plane wave solutions begins to look questionable.\footnote{Recall the response to option (ii) in \S\ref{signalspeed}.} 

In response to this, one might argue that such abnormal light propagation would only arise in cosmological scenarios as the result of finely-tuned set-ups which are unlikely to occur naturally. As was pointed out earlier, one of the central conditions for the occurrence of the abnormal light propagation of Airy-type is that there is an Airy-wave producing filter that is sufficiently larger in extension than the actual beam circumference. At this stage, we can only speculate as to what kind of physical structure can give rise to such an expansive filter in outer space; however, the point remains that if one detects a beam from outer space it cannot necessarily be expected to arise as a straight signal from a single source; rather, one can \emph{prima facie} also hypothesise it to be bent and originating from an (effective) larger-scale aperture.
{\color{black} This is a crucial inquiry, as
 astrophysical sources can emit light signals with different wavepacket properties---this is evident, since both light pulses \cite{AlonsoHernández1,Jenet1f,Everett1q,McLaughlinqq,Andrewwww,DiPang} and light beams with orbital angular momentum \cite{anzolin,Harwit,Berkhout1} have been observed and studied.}

In the remainder of this section, we explore the conceptual significance of the (assumed) existence of such non-standard electromagnetic wave solutions in astrophysical scenarios with respect both to the underdetermination of theory by data (\S\ref{s6.1}), and with respect to the theory-dependence of observation (\S\ref{s6.2}).

\subsection{Underdetermination of theory by data}\label{s6.1}

Underdetermination of theory by data is often posed in the philosophy of science in a `global' manner: if two different theories make the same empirical predictions, then how is one to tell which of those theory `truly' describes reality? Issues of underdetermination, however, can of course also be a `local' matter:\footnote{What we mean here is essentially tracking the difference between `strong' versus `weak' underdetermination, as summarised in e.g.~\cite{Ladyman2001-LADUPO-2}.} given such-and-such empirical data/evidence, which theoretical hypotheses are compatible with that data/evidence?\footnote{In what follows, we elide some philosophical subtleties by using `data' and `evidence' interchangeably.}

Clearly, it is this latter notion of underdetermination of theory by data which is relevant to our points in this article. In particular, given the possibility of abnormal wave 
propagation in curved spacetimes as explored in the foregoing sections of this article, there arises a novel issue of underdetermination by data \emph{vis-\`{a}-vis} the astronomical electromagnetic signals which we receive on (or around, via satellites) the Earth. In particular---and this in addition to the fact that the trajectories of photons are well-known to bend in light of the curvature of spacetime in general relativity---:
\begin{enumerate}
    \item Are the detected signals truly propagating on null geodesics and with determinate velocity $c$?
    \item Are the detected signals in fact encoding information about non-local sources?
\end{enumerate}
Given that such signals are in the relevant astrophysics and cosmology communities almost invariably \emph{assumed} to propagate locally as plane waves on null geodesics, there is---in light of the above---a pressing need to confront these questions.

In particular, in an astrophysical context, the roughly-detected beam segments as we register them on or around the Earth seem to leave open whether they are part of standard beam originating in sources `straight ahead', or are better fitted to a model of abnormal wave propagation that is associated to a  different local source (or maybe even to various sources at once). 

This issue of underdetermination arising from the possibility of abnormal wave propagation is reminiscent of how gravitational lensing for light was traditionally regarded as a source for underdetermination by data of astrophysical phenomena. There is, for instance, the concrete suggestion by Barnothy and Barnothy from 1965 \cite{Barnothy} that quasars are Seyfert galaxies subject to gravitational lensing rather than a class of cosmic objects of their own. Notably though, by now one can fairly well decide between whether or not gravitational lenses have influenced a beam's path---the theory behind lensing has evolved and given rise to decisive observational traces of lensing in the beam \cite{Anderl}. 

The currently-presented issue of underdetermination might be resolved in a similar manner. In this context, it is interesting to note the fair amount of optimism in fields of `historical' research (to which astrophysics and cosmology can be counted to the extent that they make sense of past data) to the effect that apparent issues of underdetermination by data can eventually be resolved (and thus are in particular not issues of in-principle underdetermination). Consider, for example, the following passage from \cite{Anderl} (echoing \cite{cleland}):\footnote{Cf.~the optimism about the historical sciences defended by Currie \cite{Currie}. Here is Cleland on this issue:
\begin{quote}
Historical researchers investigating particular past events cannot test their hypotheses by performing controlled experiments. But this doesn’t mean that they cannot procure empirical evidence for them. Because of the asymmetry of over-determination, there are usually an enormous number of subcollections of the effects of a past event that are individually sufficient (given the right theoretical assumptions) to infer its occurrence. The trick is finding them. Many of these overdetermining traces (e.g.,~a splinter of glass, a footprint) occupy small, local regions of space, bringing them within the limited range of human sensory experience. This places scientists investigating the remote past in the position of criminal investigators. Just as there are many different possibilities for catching a criminal, so there are many different possibilities for establishing what caused the demise of the dinosaurs, the origin of the universe, etc. Like criminal investigators, historical scientists collect evidence, consider different suspects, and follow up leads. (p.~490) \end{quote}}
\begin{quote}
[I]f an event has occurred, it is very difficult to make things look like nothing had happened. The situation seems analogous to criminal cases, where it is reasonable to hope that the culprit has left some traces that will make it possible to identify him. In this sense, it is difficult to fake past events, i.e. to create all expectable traces artificially without the occurrence of the actual event. While in an experimental approach, the underdetermination problem makes it difficult to find the ``true'' (responsible) causes for an observed outcome, the richness of traces in historical science cases can help to distinguish true causes from faked causes. The large number of effects that an event usually creates leads to a so called ``overdetermination of causes''. (p.~6) 
\end{quote}

So far, so good---but what about now? How strong really is the case of underdetermination raised by abnormal signal propagation? Are there genuine, concrete cases in which explicit alternative hypotheses for an arriving {part of the} wave package in terms of non-geodesic (possibly even non-localised) sources are known or can be found?

Let us endeavour to make some progress on this complex of questions by comparing the situation of beam detection in astrophysics to that in a lab. The difference between beam detection in astrophysics and in a standard lab scenario is striking: while in the latter case the full extent of the beam can be studied from various perspectives (using all sorts of manipulations and measurement across the full length of the beam), in the former case the beam path is only known around its detection point (which is, given the distance the beam usually has travelled in an astrophysical context, a rather insignificant segment of the overall beam path). 
Therefore---much more than in the laboratory---in astrophysics theoretical background assumptions are required for making a statement on the full beam path. It is here that the results on abnormal light propagation in curved spacetime come in: they call into question the background assumption that detected parts of electromagnetic waves follow null geodesics of the background geometry; more than that: concrete alternatives for long-term beam propagation are provided (say in the form of specifically bent Airy wave packages), albeit only with respect to beam's \textit{path}.



Eventually, however, one should be able to decide by crucial experiment between specific abnormal Airy wave solutions and standard (geodesic) beam solutions: the characteristic mark of the abnormal Airy wave solution is a specific polarisation of light that should be subject to detection, even locally.\footnote{On `crucial experiments' more generally, see \cite{Lakatos}.} Likewise, it is to be expected that other forms of abnormal light propagation will have signatures that can ultimately make them distinguishable from (say) Gaussian beams.

\subsection{Theory-dependence of observation}\label{s6.2}

Given the assumption that the detected part of light propagates on null geodesics which is at least implicit (and, indeed, sometimes explicit) in most reasoning in astrophysics and cosmology, those physicists' descriptions of observed results inevitably become a function of that assumption. For example, we hear regularly---on the basis of the assumption---that such-and-such galaxy is such-and-such many millions of light years away. Although such statements might appear to be matters of brute empirical fact, the above discussion should make clear that they are in fact much more theory-laden than they might initially appear.

Clearly, the central point of the present article is that we should be open to revising assumptions of this kind---if, on the basis of more detailed considerations (say, to do with how said electromagnetic waves are generated in such-and-such astronomical body, e.g.~a star, or filtered in such-and-such nebula), we come to appreciate that we are not dealing with plane waves, and that (rather) the signals under consideration need not manifest geodesic motion, or need not encode only local effects, then in that case our descriptions of what we are observing on the basis of received signals will change. It seems to us correct to state that a process of reflective equilibrium is an appropriate way to characterise this process: (i) astrophysics (in this case) begins with certain general and univocal assumptions (e.g.,~to do with the null propagation of light), then (ii) uses those assumptions to gather data on astrophysical bodies (e.g.~stars) and to develop a more sophisticated physics of the behaviour of those bodies, but (iii) that very understanding might lead us to revise the very assumptions from which we began (in this case, to do with the propagation of light emitted from such bodies). In turn, this only goes to illustrate the complex interrelations between the theoretical and the empirical in modern science; the lesson, of course, is that we should not (and need not) be too keen to hold onto established consensus, if theoretical considerations on the basis of amassed empirical evidence (which might have been gathered from within the framework of that very consensus) speak against it.\footnote{For a general introduction to notions of reflective equilibrium, see \cite{sep-reflective-equilibrium}. What we have in mind in this article is akin (we take it) to the iterative revision of the concept of temperature as discussed in \cite{Chang2004-CHAITM}, but in this case the target phenomena are (of course) electromagnetic signals.}

\section*{Author declarations}

No conflicts of interest arose in the creation of this work.

\section*{Acknowledgements}

We are grateful to Zhigang Chen, Patrick D\"{u}rr, and Nikos Efremidis for helpful discussions. We are also grateful to the two anonymous referees for helpful feedback and critical scrutiny. J.R.~is supported by a Leverhulme Trust Research Fellowship titled `Measuring Spacetime'. N.L. acknowledges the support of the Swiss National Science Foundation as part of the project `Philosophy Beyond Standard Physics' (105212\_207951).

\bibliographystyle{alpha}
\bibliography{main}

\newcommand{\etalchar}[1]{$^{#1}$}
\begin{thebibliography}{AMCK{\etalchar{+}}16}

\bibitem[AH17a]{AHbire}
F.~A. Asenjo and S.~A. Hojman.
\newblock Birefringent light propagation on anisotropic cosmological
  backgrounds.
\newblock {\em Phys. Rev. D}, 99:044021, 2017.

\bibitem[AH17b]{AH}
F.~A. Asenjo and S.~A. Hojman.
\newblock Do electromagnetic waves always propagate along null geodesics?
\newblock {\em Classical and Quantum Gravity}, 34(20):205011, sep 2017.

\bibitem[AHFK{\etalchar{+}}22]{AlonsoHernández1}
J.~Alonso-Hern\'andez, F.~F\"urst, P.~Kretschmar, I.~Caballero, and A.~M.
  Joyce.
\newblock Common patterns in pulse profiles of high-mass x-ray binaries.
\newblock {\em A\&A}, 662:A62, 2022.

\bibitem[AJ61]{jahnG}
F.~A. Albini and R.~G. Jahn.
\newblock Reflection and transmission of electromagnetic waves at electron
  density gradients.
\newblock {\em J. Appl. Phys.}, 32:75, 1961.

\bibitem[AMCK{\etalchar{+}}16]{Parinaz2016}
P.~Aleahmad, H.~Moya~Cessa, I.~Kaminer, M.~Segev, and Christodoulides~D. N.
\newblock Dynamics of accelerating {B}essel solutions of {M}axwell’s
  equations.
\newblock {\em Journal of the Optical Society of America A}, 33(10):2047, 2016.

\bibitem[And15]{Anderl}
Sibylle Anderl.
\newblock Astronomy and astrophysics in the philosophy of science.
\newblock {\em arXiv preprint arXiv:1510.03284}, 2015.

\bibitem[ATB{\etalchar{+}}08]{anzolin}
G.~Anzolin, F.~Tamburini, A.~Bianchini1, G.~Umbriaco, and C.~Barbieri.
\newblock Optical vortices with starlight.
\newblock {\em A\&A}, 488:1159, 2008.

\bibitem[BB68]{Barnothy}
J.~M. Barnothy and M.~F. Barnothy.
\newblock The relationship between {S}eyfert galaxies, $n$-type galaxies and
  quasi-stellar objects.
\newblock {\em Astrophysical Letters}, 2:21--26, 1968.

\bibitem[BB79]{berry1979}
M.~V. Berry and N.~L. Balazs.
\newblock Nonspreading wave packets.
\newblock {\em American Journal of Physics}, 47(3):264, 1979.

\bibitem[BB09]{Berkhout1}
G.~C.~G. Berkhout and M.~W. Beijersbergen.
\newblock Using a multipoint interferometer to measure the orbital angular
  momentum of light in astrophysics.
\newblock {\em J. Opt. A: Pure Appl. Opt.}, 11:094021, 2009.

\bibitem[BNKS14]{bekenstein2014shape}
Rivka Bekenstein, Jonathan Nemirovsky, Ido Kaminer, and Mordechai Segev.
\newblock Shape-preserving accelerating electromagnetic wave packets in curved
  space.
\newblock {\em Physical Review X}, 4(1):011038, 2014.

\bibitem[BSDC08a]{broky2008}
John Broky, Georgios~A Siviloglou, Aristide Dogariu, and Demetrios~N
  Christodoulides.
\newblock Self-healing properties of optical airy beams.
\newblock {\em Optics express}, 16(17):12880--12891, 2008.

\bibitem[BSDC08b]{broky2008self}
John Broky, Georgios~A Siviloglou, Aristide Dogariu, and Demetrios~N
  Christodoulides.
\newblock Self-healing properties of optical airy beams.
\newblock {\em Optics express}, 16(17):12880--12891, 2008.

\bibitem[But07]{Butterfield}
Jeremy Butterfield.
\newblock Reconsidering relativistic causality.
\newblock {\em International Studies in the Philosophy of Science},
  21(3):295--328, 2007.

\bibitem[Cao01]{Cao}
Tian~Yu Cao.
\newblock Prerequisites for a consistent framework of quantum gravity.
\newblock {\em Studies in History and Philosophy of Science Part B: Studies in
  History and Philosophy of Modern Physics}, 32(2):181--204, 2001.
\newblock Spacetime, Fields and Understanding: Persepectives on Quantum Field.

\bibitem[CE13]{Chremmos2013}
I.~D. Chremmos and N.~K. Efremidis.
\newblock Nonparaxial accelerating {B}essel-like beams.
\newblock {\em Phys. Rev. A}, 88:063816, 2013.

\bibitem[Cha04]{Chang2004-CHAITM}
Hasok Chang.
\newblock {\em Inventing Temperature: Measurement and Scientific Progress}.
\newblock New York, US: OUP Usa, 2004.

\bibitem[Cle02]{cleland}
Carol~E Cleland.
\newblock Methodological and epistemic differences between historical science
  and experimental science.
\newblock {\em Philosophy of science}, 69(3):474--496, 2002.

\bibitem[CPS21]{lightCP3}
C.~J. Copi, K.~Pasmatsiou, and G.~D. Starkman.
\newblock Scalar and vector tail radiation from the interior of the lightcone.
\newblock {\em JCAP}, 50, 2021.

\bibitem[CRNW10]{chong2010}
A.~Chong, W.~H. Renninger, Christodoulides~D. N., and Wise~F. W.
\newblock Airy–{B}essel wave packets as versatile linear light bullets.
\newblock {\em Naure Photonics}, 4:103, 2010.

\bibitem[Cur18]{Currie}
Adrian Currie.
\newblock {\em Rock, Bone, and Ruin: An Optimist's Guide to the Historical
  Sciences}.
\newblock Life and Mind: Philosophical Issues in Biology and Psychology. MIT
  Press, 2018.

\bibitem[Dah23]{lastnewf3}
P.~K. Dahal.
\newblock Light propagation in kerr spacetime.
\newblock {\em Eur. Phys. J. Plus}, 138:205, 2023.

\bibitem[Dan20]{sep-reflective-equilibrium}
Norman Daniels.
\newblock {Reflective Equilibrium}.
\newblock In Edward~N. Zalta, editor, {\em The {Stanford} Encyclopedia of
  Philosophy}. Metaphysics Research Lab, Stanford University, {S}ummer 2020
  edition, 2020.

\bibitem[DB60]{lightCP1}
B.~S. DeWitt and R.~W. Brehme.
\newblock Radiation damping in a gravitational field.
\newblock {\em Annals of Physics}, 9:220, 1960.

\bibitem[ECSC19]{Airyreview}
Nikolaos~K Efremidis, Zhigang Chen, Mordechai Segev, and Demetrios~N
  Christodoulides.
\newblock Airy beams and accelerating waves: an overview of recent advances.
\newblock {\em Optica}, 6(5):686--701, 2019.

\bibitem[{Ein}05]{Einstein1905}
A.~{Einstein}.
\newblock {Zur Elektrodynamik bewegter K{\"o}rper}.
\newblock {\em Annalen der Physik}, 322(10):891--921, January 1905.

\bibitem[Ein54]{Einstein1954}
Albert Einstein.
\newblock {\em Ideas and Opinions}.
\newblock Bonanza Books, 1954.

\bibitem[EW01]{Everett1q}
J.~E. Everett and J.~M. Weisberg.
\newblock Emission beam geometry of selected pulsars derived from average pulse
  polarization data.
\newblock {\em ApJ}, 553:341, 2001.

\bibitem[GBM07]{lastnewf1}
P.~Gosselin, A.~B\'erard, and H.~Mohrbach.
\newblock Spin hall effect of photons in a static gravitational field.
\newblock {\em Phys. Rev. D}, 75:084035, 2007.

\bibitem[GMCB98]{GARRISON199819}
J.C. Garrison, M.W. Mitchell, R.Y. Chiao, and E.L. Bolda.
\newblock Superluminal signals: causal loop paradoxes revisited.
\newblock {\em Physics Letters A}, 245(1):19--25, 1998.

\bibitem[Hac11]{Hacyan2011}
S.~Hacyan.
\newblock Relativistic accelerating electromagnetic waves.
\newblock {\em Journal of Optics}, 13:105710, 2011.

\bibitem[Har03]{Harwit}
M.~Harwit.
\newblock Photon orbital angular momentum in astrophysics.
\newblock {\em ApJ}, 597:1266, 2003.

\bibitem[Har19]{lastnewf4}
A.~I. Harte.
\newblock Gravitational lensing beyond geometric optics: I. formalism and
  observables.
\newblock {\em Gen. Rel. Grav.}, 14:205, 2019.

\bibitem[HHW{\etalchar{+}}04]{Andrewwww}
A.~W. Howard, P.~Horowitz, D.~T. Wilkinson, C.~M. Coldwell, E.~J. Groth,
  N.~Jarosik, D.~W. Latham, R.~P. Stefanik, Jr. A.~J. Willman, J.~Wolff, and
  J.~M. Zajac.
\newblock Search for nanosecond optical pulses from nearby solar-type stars.
\newblock {\em ApJ}, 613:1270, 2004.

\bibitem[HS98]{lightCP2}
B.~L. Hu and K.~Shiokawa.
\newblock Wave propagation in stochastic spacetimes: Localization,
  amplification, and particle creation.
\newblock {\em Phys. Rev. D}, 57:3474, 1998.

\bibitem[JFF{\etalchar{+}}10]{Jenet1f}
F.~A. Jenet, D.~Fleckenstein, A.~Ford, A.~Garcia, R.~Miller, J.~Rivera, and
  K.~Stovall.
\newblock Apparent faster-than-light pulse propagation in interstellar space: A
  new probe of the interstellar medium.
\newblock {\em ApJ}, 710:1718, 2010.

\bibitem[KBNS12]{beenstein2012prl}
Ido Kaminer, Rivka Bekenstein, Jonathan Nemirovsky, and Mordechai Segev.
\newblock Nondiffracting accelerating wave packets of {M}axwell’s equations.
\newblock {\em Physical Review Letters}, 108:163901, 2012.

\bibitem[KH10]{Kaganovsky}
Y~Kaganovsky and E~Heyman.
\newblock Wave analysis of airy beams.
\newblock {\em Optics Express}, 18(8):8440--8452, 2010.

\bibitem[Lad01]{Ladyman2001-LADUPO-2}
James Ladyman.
\newblock {\em Understanding Philosophy of Science}.
\newblock Routledge, 2001.

\bibitem[Lak74]{Lakatos}
Imre Lakatos.
\newblock The role of crucial experiments in science.
\newblock {\em Studies in History and Philosophy of Science Part A},
  4(4):309--325, 1974.

\bibitem[Lek04]{lekner2003}
J.~Lekner.
\newblock Energy and momentum of electromagnetic pulses.
\newblock {\em J. Opt. A: Pure Appl. Opt.}, 6:146, 2004.

\bibitem[Lek09]{lekner2009}
J.~Lekner.
\newblock Airy wavepacket solutions of the schr\"odinger equation.
\newblock {\em Eur. J. Phys.}, 30:L43, 2009.

\bibitem[LR21]{LR}
Niels Linnemann and James Read.
\newblock Comment on `{D}o electromagnetic waves always propagate along null
  geodesics?'.
\newblock {\em Classical and Quantum Gravity}, 38(23):238001, dec 2021.

\bibitem[Man07]{Mannheim2007}
Philip~D. Mannheim.
\newblock Solution to the ghost problem in fourth order derivative theories.
\newblock {\em Foundations of Physics}, 37(4):532--571, 2007.

\bibitem[Man22]{lightCP4}
P.~D. Mannheim.
\newblock Critique of the use of geodesics in astrophysics and cosmology.
\newblock {\em Class. Quantum Grav.}, 39:245001, 2022.

\bibitem[MC03]{McLaughlinqq}
M.~A. McLaughlin and J.~M. Cordes.
\newblock Searches for giant pulses from extragalactic pulsars.
\newblock {\em ApJ}, 596:982, 2003.

\bibitem[MTW73]{MTW}
Charles~W. Misner, K.~S. Thorne, and J.~A. Wheeler.
\newblock {\em {Gravitation}}.
\newblock W. H. Freeman, San Francisco, 1973.

\bibitem[NEC{\etalchar{+}}18]{Nichols}
J.~Nichols, Tegan Emerson, L.~Cattell, Serim Park, Andrey Kanaev, Frank
  Bucholtz, Abbie Watnik, T.~Doster, and G.~Rohde.
\newblock Transport-based model for turbulence-corrupted imagery.
\newblock {\em Applied Optics}, 57:4524, 06 2018.

\bibitem[NNB22]{Nichols:22}
J.~M. Nichols, D.~V. Nickel, and F.~Bucholtz.
\newblock Vector beam bending via a polarization gradient.
\newblock {\em Opt. Express}, 30(21):38907--38929, Oct 2022.

\bibitem[PBBS18]{patsyk2018observation}
Anatoly Patsyk, Miguel~A Bandres, Rivka Bekenstein, and Mordechai Segev.
\newblock Observation of accelerating wave packets in curved space.
\newblock {\em Physical Review X}, 8(1):011001, 2018.

\bibitem[PCP{\etalchar{+}}16]{peng2016}
P.~Peng, B.~Chen, X.~Peng, M.~Zhou, L.~Zhang, D.~Li, and D.~Deng.
\newblock Self-accelerating airy-ince-gaussian and airy-helical-ince-gaussian
  light bullets in free space.
\newblock {\em Optics Express}, 24(17):18973, 2016.

\bibitem[PGPDM18]{DiPang}
D.~Pang, K.~Goseva-Popstojanova, T.~Devine, and M.~McLaughlin.
\newblock A novel single-pulse search approach to detection of dispersed radio
  pulses using clustering and supervised machine learning.
\newblock {\em Month. Not. Roy. Astr. Soc.}, 480:3302, 2018.

\bibitem[PN98]{PRL}
D.~Paganin and K.~A. Nugent.
\newblock Noninterferometric phase imaging with partially coherent light.
\newblock {\em Phys. Rev. Lett.}, 80:2586--2589, Mar 1998.

\bibitem[PSAT10]{papazoglou2010}
D.~G. Papazoglou, S.~Suntsov, D.~Abdollahpour, and S.~Tzortzakis.
\newblock Tunable intense airy beams and tailored femtosecond laser filaments.
\newblock {\em Phys. Rev. A}, 81:061807, Jun 2010.

\bibitem[PTB13]{Petruccelli:13}
Jonathan~C. Petruccelli, Lei Tian, and George Barbastathis.
\newblock The transport of intensity equation for optical path length recovery
  using partially coherent illumination.
\newblock {\em Opt. Express}, 21(12):14430--14441, Jun 2013.

\bibitem[SBDC07]{siviloglou2007}
G.~A. Siviloglou, J.~Broky, A.~Dogariu, and D.~N. Christodoulides.
\newblock Observation of accelerating airy beams.
\newblock {\em Phys. Rev. Lett.}, 99:213901, 2007.

\bibitem[SBDC08]{siviloglou2008}
G.~A. Siviloglou, J.~Broky, A.~Dogariu, and D.~N. Christodoulides.
\newblock Ballistic dynamics of airy beams.
\newblock {\em Opt. Lett.}, 33(3):207--209, Feb 2008.

\bibitem[SC07]{Siviloglou2007a}
G.~A. Siviloglou and D.~N. Christodoulides.
\newblock Accelerating finite energy airy beams.
\newblock {\em Optics Letters}, 32:979, 2007.

\bibitem[SGN03]{Stenner}
M.~D. Stenner, D.~J. Gauthier, and M.~A. Neifeld.
\newblock The speed of information in a `fast-light'optical medium.
\newblock {\em Nature}, 425(6959):695--698, 2003.

\bibitem[Sho03]{Shore}
G.~M. Shore.
\newblock {Causality and superluminal light}.
\newblock In {\em {Time and Matter}: {An International Colloquium on the
  Science of Time}}, pages 45--66, 2 2003.

\bibitem[SSTL23]{TellezLimon}
C.~T. Sosa-S\'anchez and R.~T\'ellez-Lim\'on.
\newblock Plasmonic metalens to generate an airy beam.
\newblock {\em Nanomaterials}, 13:2576, 2023.

\bibitem[TA10]{tetryakov}
O.~A. Tretyakov and O.~Akgun.
\newblock Derivation of klein-gordon equation from maxwell's equations and
  study of relativistic time-domain waveguides modes.
\newblock {\em Prog. Electromag. Res.}, 105:171, 2010.

\bibitem[vdBF20]{lastnewf2}
P.~M. van~den Berg and J.~T. Fokkema.
\newblock New insight in the propagation of wave fronts in the vicinity of a
  refractive object.
\newblock {\em URSI Radio Science Letters}, 2:1, 2020.

\bibitem[YSA10]{yalizay2010}
Berna Yalizay, Burak Soylu, and Selcuk Akturk.
\newblock Optical element for generation of accelerating airy beams.
\newblock {\em J. Opt. Soc. Am. A}, 27(10):2344--2346, Oct 2010.

\bibitem[ZBH13]{ping2013}
W.-P. Zhong, M.~Beli\'c, and T.~Huang.
\newblock Three-dimensional finite-energy airy self-accelerating
  parabolic-cylinder light bullets.
\newblock {\em Phys. Rev. A}, 88:033824, 2013.

\bibitem[ZZMS15]{zhuang2015quantitative}
Fei Zhuang, Ziyi Zhu, Jessica Margiewicz, and Zhimin Shi.
\newblock Quantitative study on propagation and healing of airy beams under
  experimental conditions.
\newblock {\em Optics Letters}, 40(5):780--783, 2015.

\end{thebibliography}

\end{document}